\documentclass{PoS}
\pdfoutput=1
\title{Progress of the MICE experiment}

\ShortTitle{Progress of the MICE experiment}
\author{\speaker{Maurizio Bonesini}%
         \thanks{on behalf of the MICE Collaboration}\\
        Sezione INFN Milano-Bicocca, Dipartimento di Fisica G. Occhialini, \\
        Piazza Scienza 3, 20123 Milano, Italy\\
        E-mail: \email{maurizio.bonesini@mib.ch}}

\abstract{
The international Muon Ionization Cooling Experiment (MICE) will perform a 
systematic investigation of ionization cooling of a muon beam.
The demonstration is based on  a simplified version of 
a neutrino factory 
cooling channel.
As the emittance measurement will be done on a particle-by-particle
basis, sophisticated beam instrumentation has been developed to measure
particle coordinates and timing vs RF. 
The muon beamline has been characterized and a
preliminary measure of the beam emittance, using a particle-by-particle
method with only the TOF detector system, has been performed (MICE 
STEP I).  Data taking for the study of the properties
that determine the cooling performance (MICE Step IV)  has just started in 2015,
while the demonstration of ionization cooling with re-acceleration is foreseen
for 2017. 
}

\FullConference{The European Physical Society Conference on High Energy Physics\\
		 22-29 July 2015\\
		 Vienna, Austria}

\begin{document}

\section{Introduction}
A neutrino factory~\cite{Koshkarev} is a muon
storage ring with long straight
sections, where decaying muons produce collimated, high intensity 
neutrino beams of defined
composition ($50 \% \ \nu_{e}, 50 \% \ \overline{\nu}_{\mu}$ for
the $\mu^{+} \rightarrow \overline{\nu}_{\mu} \nu_e e^{+}$ case), 
with no uncertainties in the spectrum and flux
from hadronic production~\cite{Bonesini}.
A neutrino factory will be the most efficient tool to probe the neutrino
sector and observe CP violation in leptons.
In addition, at the high energy frontier, colliding muon
beams may be a valuable option, benefitting from the use of
point-like particles 
and the much higher mass of a muon with respect to that of 
the electron~\cite{Ankenbrandt}.

The cooling of muons
(accounting for $ \sim 20 \%$ of the final costs)
will increase the neutrino factory performance and reduce the
muon beam emittance up to a factor 2.4
(as described in \cite{Choubey} with a cooling section 75 m long).

The  modest muon cooling needs of a neutrino factory
might be traded off by using a larger aperture machine, such as
a Fixed-Field Alternating Gradient accelerator~\cite{Rees}, 
but no practical muon collider \cite{Geer1} is
conceivable  without cooling of at least three orders of magnitude.

Conventional beam cooling methods are ineffective on the short 
timescale of muon lifetime ($\tau \sim 2.2 \mu$s). 
The only effective way is the 
so-called ``ionization cooling'' that is accomplished by passing
muons through a low-Z absorber, where they loose energy by 
ionization 
and the longitudinal component of momentum is then
replenished by RF cavities \cite{srinsky}. 

\begin{figure}[thb]
\begin{center}
\includegraphics[width=\linewidth]{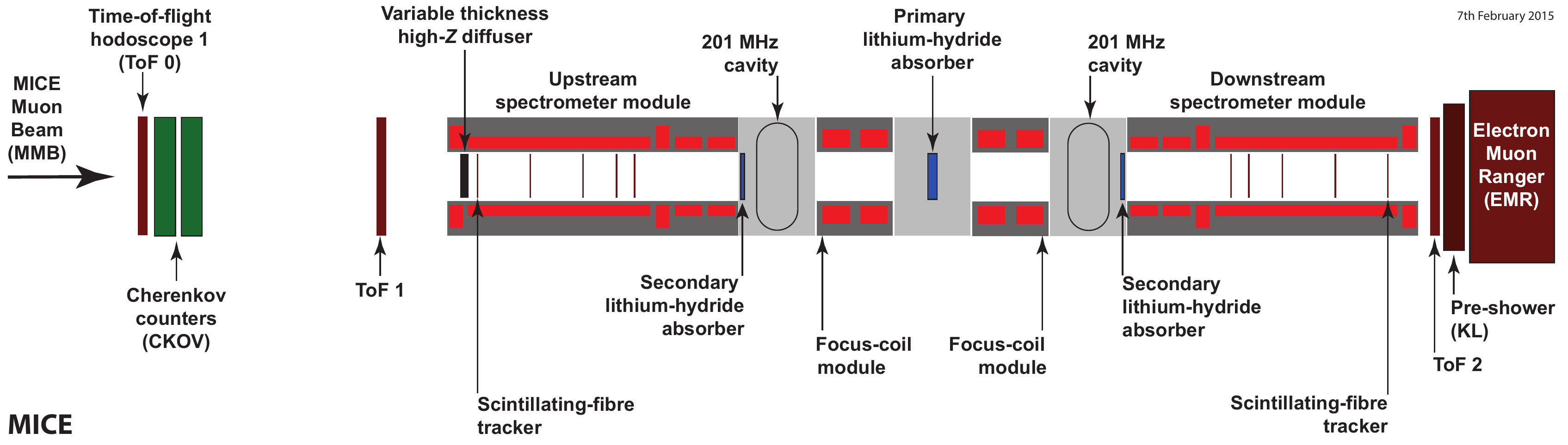}
\label{fig-mice}
\end{center}
\vskip -1cm
\caption{View of the MICE experiment at RAL.
The muon beam from ISIS enters from the left. The cooling channel
is put between two magnetic spectrometers and two TOF stations
(TOF1 and TOF2) to measure particle parameters.}
\end{figure}

The initial goal of the MICE experiment \cite{mice}  to study
a fully engineered cooling cell  of the 
proposed US Study 2~\cite{US2a}~\footnote{
a 5.5 m long cooling section consisting of three liquid hydrogen absorbers
and eight 201 MHz RF cavities encircled by lattice solenoids}, has been
downsized in 2014 to a demonstration of ionization cooling with a simplified
lattice based on the
available RF cavities and absorber-focus coils (see figure \ref{fig-mice}). 

MICE is being done in several steps, of which the first one (STEP I) is 
the characterization of the beamline.

\section{The MICE STEPI beamline characterization}

The dedicated muon beam from ISIS \footnote{140-240 MeV/c central momentum, 
tunable between $3-10 \pi \cdot  $ mm rad  input emittance} enters
the MICE cooling section after a Pb diffuser of adjustable thickness. 
Pions are produced dipping a hollow titanium cylindrical target into the 
ISIS proton beam and then captured by a quadrupole triplet (Q1-Q3). 
The dip depth and timing with respect to the ISIS 
beam cycle determine the production rate. 
Muons are then produced  from $\pi$
decay inside a 5 m long superconducting (SC) solenoid (DS) upstream of the first PID detectors.

\begin{figure}[hbt]
\begin{center}
\includegraphics[width=0.45\textwidth]{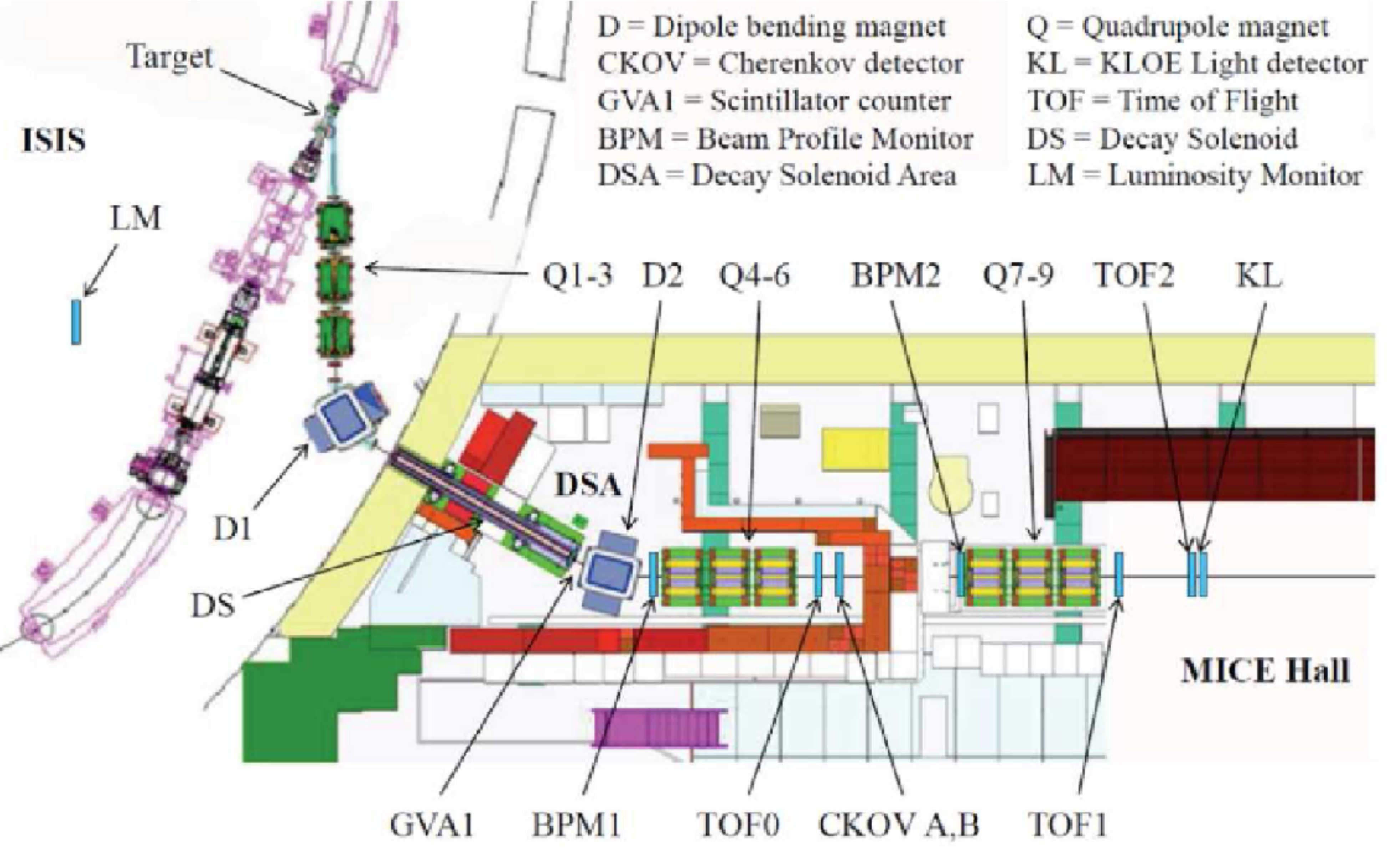}
\caption{ Sketch of the  MICE beamline, with installed detectors for
          STEPI.}
\label{fig:tofes}
\end{center}
\end{figure}

Particle identification (PID)
 is obtained upstream of the first tracking solenoid by two
TOF stations (TOF0/TOF1) \cite{yordan} and two threshold Cerenkov counters (CKOVa/CKOVb)
\cite{ckv},
that will provide $\pi/\mu$ separation up to 365 MeV/c. 
A sketch of the present MICE beamline is shown in 
figure \ref{fig:tofes}.

\noindent
Downstream the PID is obtained via an additional TOF station (TOF2)
\cite{tof2}
and a calorimeter.
All TOF detectors are used to determine the time coordinate ($t$) 
in the measurement of the emittance.

A detector resolution $\sim 50 \ $ ps is needed for TOF0, to determine the timing with respect to the RF phase to a precision of $5^{0}$ . 
To have a better than $99 \% $ rejection of pions in  the incoming 
muon beam, a resolution $\sim 100 $ ps for the TOF measurement
between TOF0 and TOF1 is needed. All these requirements imply a 
conservative request of $\sim 50-60$ ps for single TOF station resolution.
All the  TOF stations share a common design
based on fast 1" scintillator counters
along the x/y directions (to increase measurement redundancy) read at both
edges by conventional fast R4998 Hamamatsu photomultipliers~\footnote{1" linear focussed PMTs,
typical gain $G \sim 5.7 \times 10^6$ at B=0 Gauss, risetime 0.7 ns, TTS
$\sim 160$ps. Tests were done in laboratory to characterize 
them with a dedicated setup \cite{bonesini2}}.

The downstream calorimeter, made of two separate detectors: KL
and EMR,  is not intended to be used for energy
measurement. Its main goal is to separate  muons from decay 
electrons and undecayed pions.
In the MICE calorimeter, EMR determines precisely the muon 
momentum by range meadurement, while KL acts as an active pre-shower to
tag electrons  from muon decay.
It consists of
a Pb-scintillating fiber calorimeter (KL), of the KLOE type
 \cite{kloe}, with 1-mm diameter blue scintillating fibers glued between
 0.3 mm thick grooved lead plates
  followed by an electron-muon ranger (EMR), made
 of a $\sim 1 m^3$ fully sensitive segmented scintillator block \cite{lietti}.

For MICE  STEP IV, the PID detectors
have been complemented by two trackers
inside the spectrometer solenoids, before and after the cooling section.
Each tracker \cite{Ellis}
 is composed of five stations, each made of  three layers of 
$350 \ \mu$m diameter scintillating fibre doublets, read out by 
Visible Light Photon Counters (VLPCs).

The MICE beamline has been characterized during the so-called STEP I 
by the use of the TOF system,
with data taken mainly in summer 2010. 

\begin{figure}[htb]
\vskip -3cm
\begin{center}
\includegraphics[width=5cm]{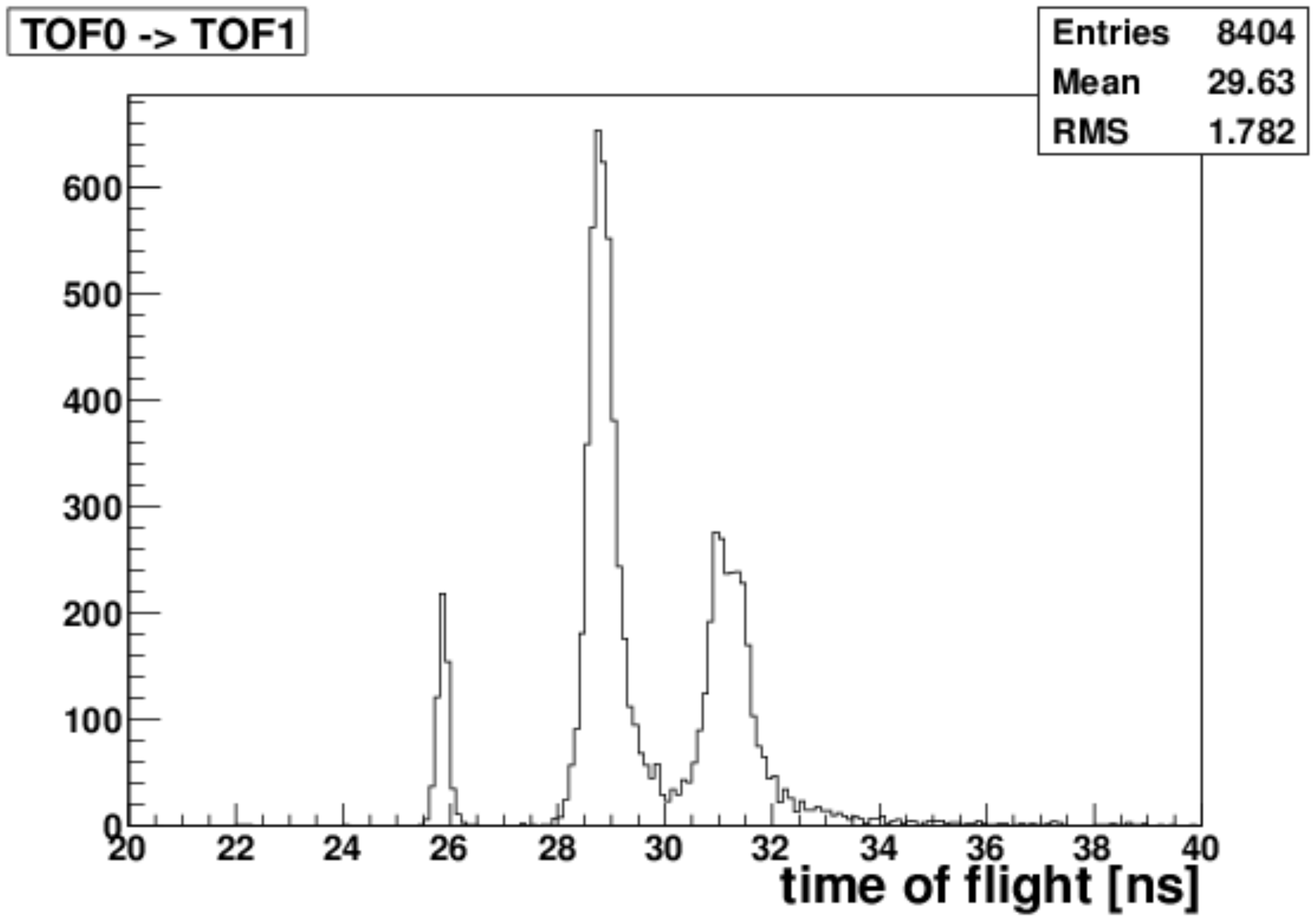}
\includegraphics[width=5cm]{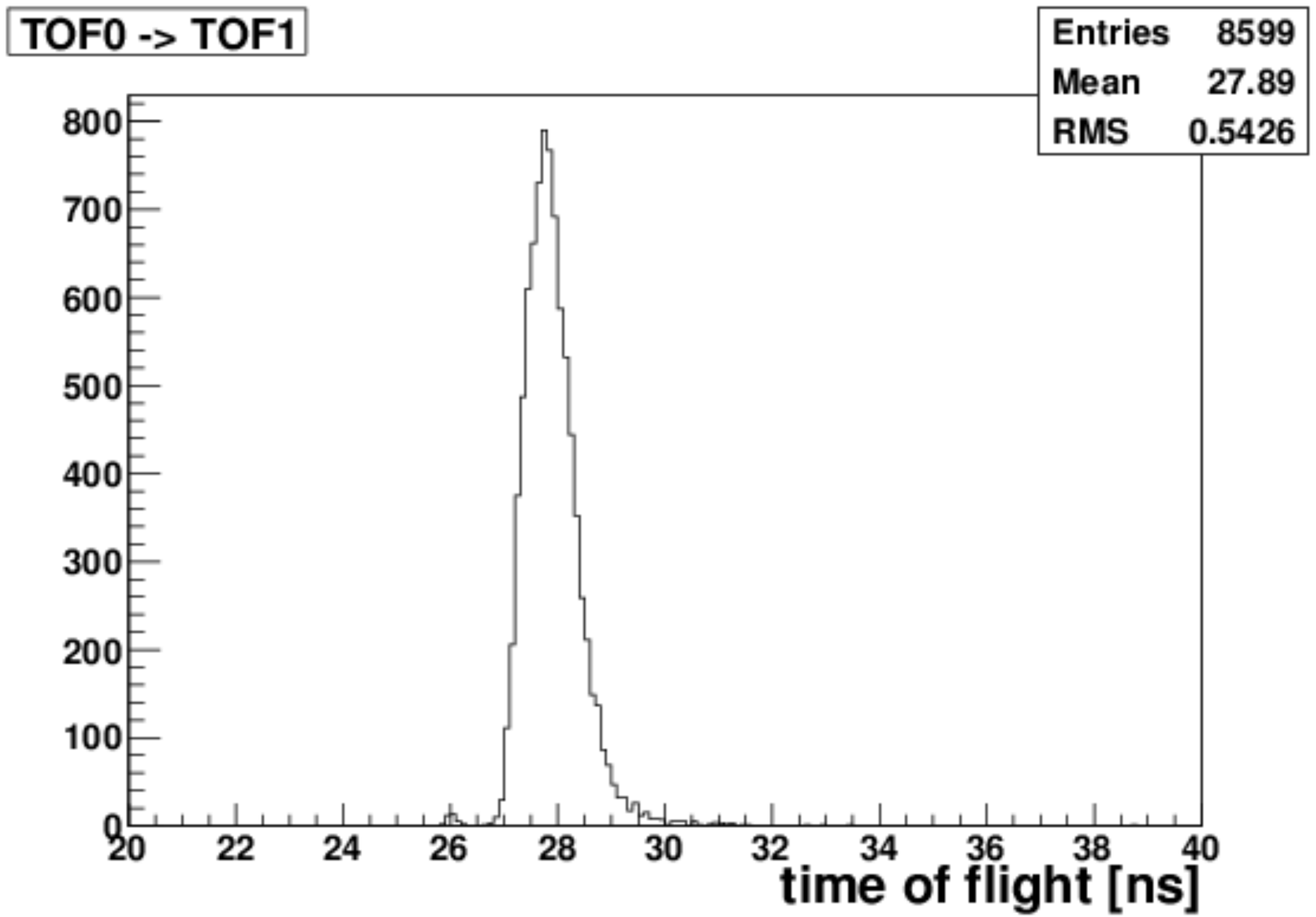}
\caption{ Time of flight between TOF0 and TOF1 for the  calibration and muon
beams.}
\label{tof}
\end{center}
\end{figure}

Figure \ref{tof} shows 
 the distribution of the
time-of-flight between TOF0 and TOF1
for a low emittance calibration beam (right panel) and a high emittance muon 
($\pi \rightarrow \mu$) beam (left panel).
The first peak which is present in both distributions 
is considered as the time-of-flight of the positrons
and is used to determine the absolute value of the time in TOF1.
A natural interpretation of the other two peaks is that they are due
to forward flying muons from pion decay and pions themselves.

 Using TOF identification, it was possible to determine the muon rate for
the $\pi \rightarrow \mu$ beam as a function of target dip into the ISIS
beam (measured as beam loss in $V \cdot ms$) \cite{bogo1}. 

As conventional emittance measurement techniques reach barely a $10 \%$ 
precision, 
the final measure of emittance will be done in the MICE beam  
on a particle-by-particle basis with  
the trackers (to measure $x,y,x'=p_x/p_z,y'=p_y/p_z.E$ for each particle) 
and the TOF stations (to measure $t$). 
In this way, for an ensemble of N ($\sim 10^6$) particles, the
input and output emittances may be  determined  with a precision
up to $1 \%$, that allows a sensible extrapolation of the results 
to the full cooling channel.
Due to a schedule delay in 
the tracking solenoids, in MICE STEPI the emittance was preliminary 
measured with the TOF system only, deriving from them also the $x,y,x',y'$
information for each particle and measuring $p_z$ from the time-of-flight
between TOF0 and TOF1 \cite{mark}.       
 
Figure 4 shows these 
distributions for the MICE baseline beam ($\epsilon=6 \pi$ mm 
and $p_z=200$ MeV/c) for experimental data and MC simulation. Even if
the agreement is not perfect, the beam occupies the desired regions in the
trace space. All beams show an RMS beam size of the order of 5-7 cm. 

\begin{figure*}[hbt!]
\vskip -2cm
\begin{center}
\includegraphics[width=\linewidth]{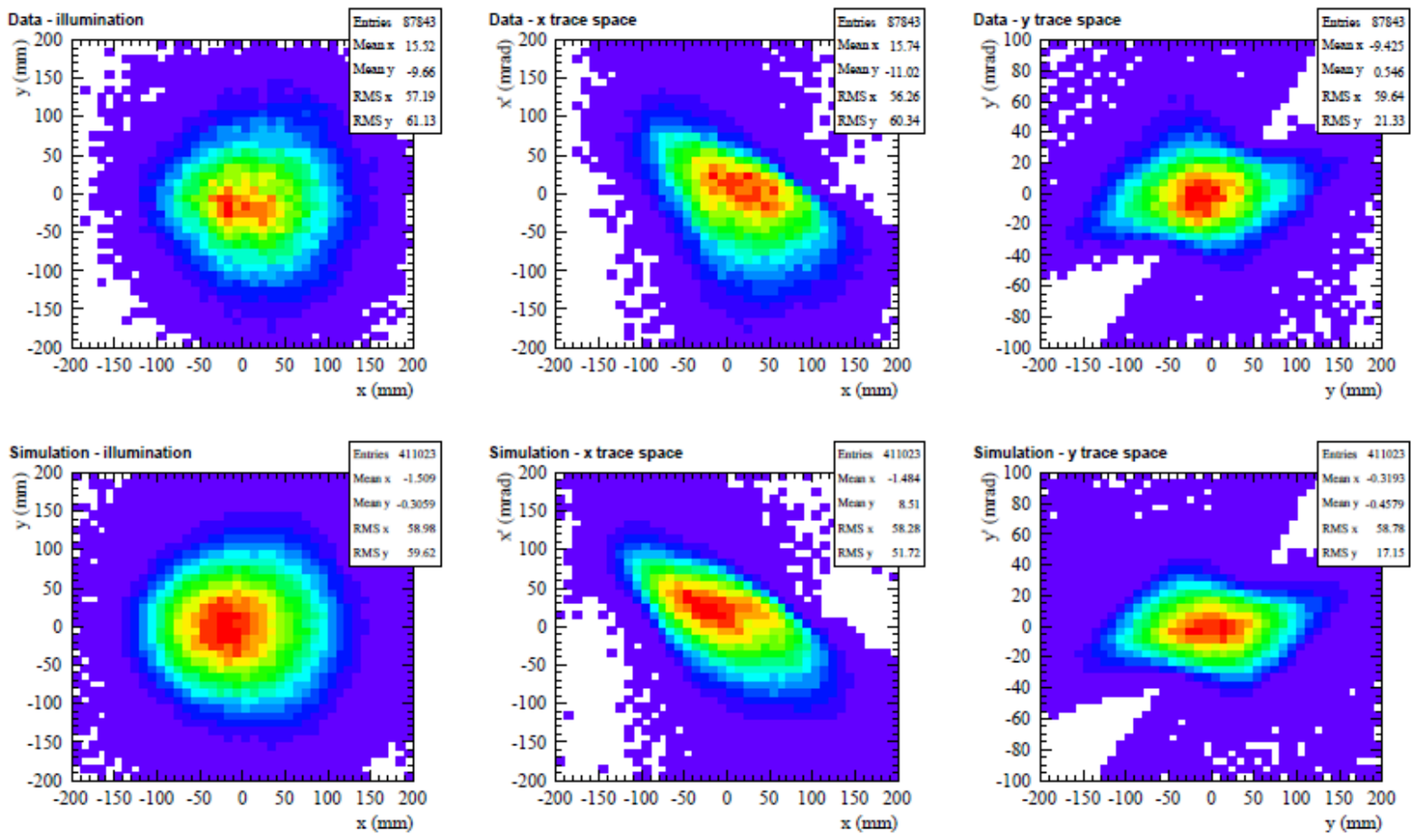}
\vskip -13cm
\caption{Reconstructed (top) and simulated (bottom) data for the trace plots
for the baseline MICE beam $\epsilon=6 \pi$ mm and $p_z=200$ MeV/c.}
\end{center}
\label{fig:trace}
\end{figure*}
In addition the pion contamination in the MICE muon beam has been measured 
to be around $1 \%$ , as requested for a high precision ($0.1 \%$)
accuracy in the emittance measurement \cite{pi-cont}. 

\section{Further MICE Steps}
A full demonstration of ionization cooling involves
a study of the properties that determine the cooling performances
 and a demonstration of transverse emittance reduction with longitudinal
re-acceleration.
The first item depends on initial beam emittance, muon momentum, absorber
material and $\beta_{\perp}$ at the absorber location and its study is the
main goal of MICE STEP IV. 
\subsection{MICE STEP IV}
\begin{figure}[thb]
\begin{center}
\includegraphics[width=\linewidth]{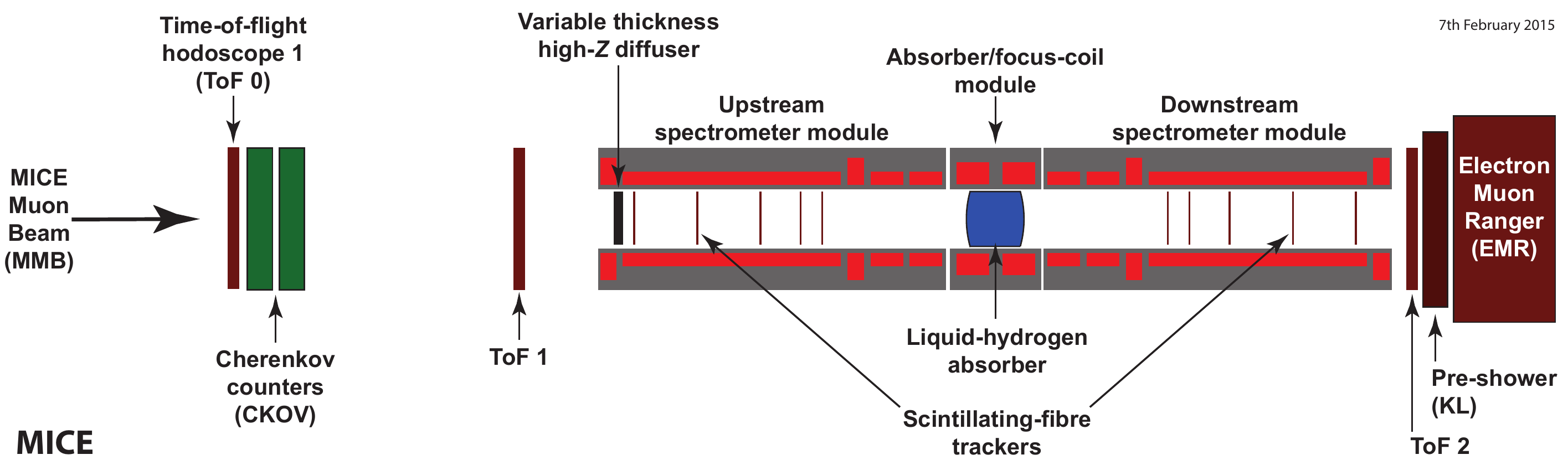}
\end{center}
\caption{View of the MICE experiment at RAL in the configuration 
to be used for STEP IV.
The muon beam from ISIS enters from the left. The absorber-focus coil
module is put between two magnetic spectrometers and two TOF stations
(TOF1 and TOF2) that measure particle parameters.}
\label{fig:mice}
\end{figure}

MICE STEP IV involves an Absorber Focus Coil (AFC) module between two
spectrometer solenoids, as shown in figure \ref{fig:mice}.
The AFC module is made of two superconducting (SC) coils surrounding the
absorber. 
For liquid hydrogen ($LH_2$) a cryogenic system is foreseen.
A list of possible absorbers is given in table \ref{tab1}.

\begin{table}[hbt]
\centering
\vspace{2mm}
\begin{tabular}{|c|c|c|c|}
\hline 
material & $ < dE/dX > _{Min} (MeV g^{-1}cm^{-2}$) & $X_0(g cm^{-2}$) &
merit factor \\ \hline
LH2 & 4.034 & 61.28 & 1 \\
He  & 1.937 & 94.32 & 0.74 \\
LiH & 1.94  & 86.9  & 0.68 \\
Li  & 1.639 & 82.76 & 0.55 \\
Be  & 1.594 & 65.19 & 0.42 \\ \hline
\end{tabular}
\label{tab1}
\caption{Comparison of possible absorber materials for ionization 
cooling.}
\end{table}

The lowest equilibrium emittance, corresponding to the optimal cooling 
channel, is obtained when $\beta_{\perp}$ is minimized and the merit 
factor ($ X_{0} \cdot \frac{dE}{dX} $) is maximized. The best merit 
factor is obtained with liquid hydrogen (a factor two better than the
next material: helium). 
MICE will take data with $LH_2$ and LiH absorbers, with the possible addition
of other materials. In addition to LiH disks, data taking with LiH wedge 
absorbers is foreseen to test longitudinal emittance reduction through 
emittance exchange \cite{bonesini1}.  

The two fiber trackers have been recently installed inside the  SC solenoids.
Each SC solenoid consists of five SC coils on a common aluminium mandrel: 
three to provide a 4T field over
the 1m long, 20 cm radius tracking volume with a better than $1 \%$ uniformity
and two to match the beam in or out the cooling cell. 
The two solenoids have been field mapped at RAL and are undergoing training;
some data 
taking for STEP IV has started.
\subsection{The cooling demonstration}
This will be the first engineering demonstration of a "sustainable
cooling", as the energy lost in absorbers will be replenished by 
RF-reacceleration. 
\begin{figure}[thb]
\vskip -0.2cm
\begin{center}
\includegraphics[width=0.35\linewidth]{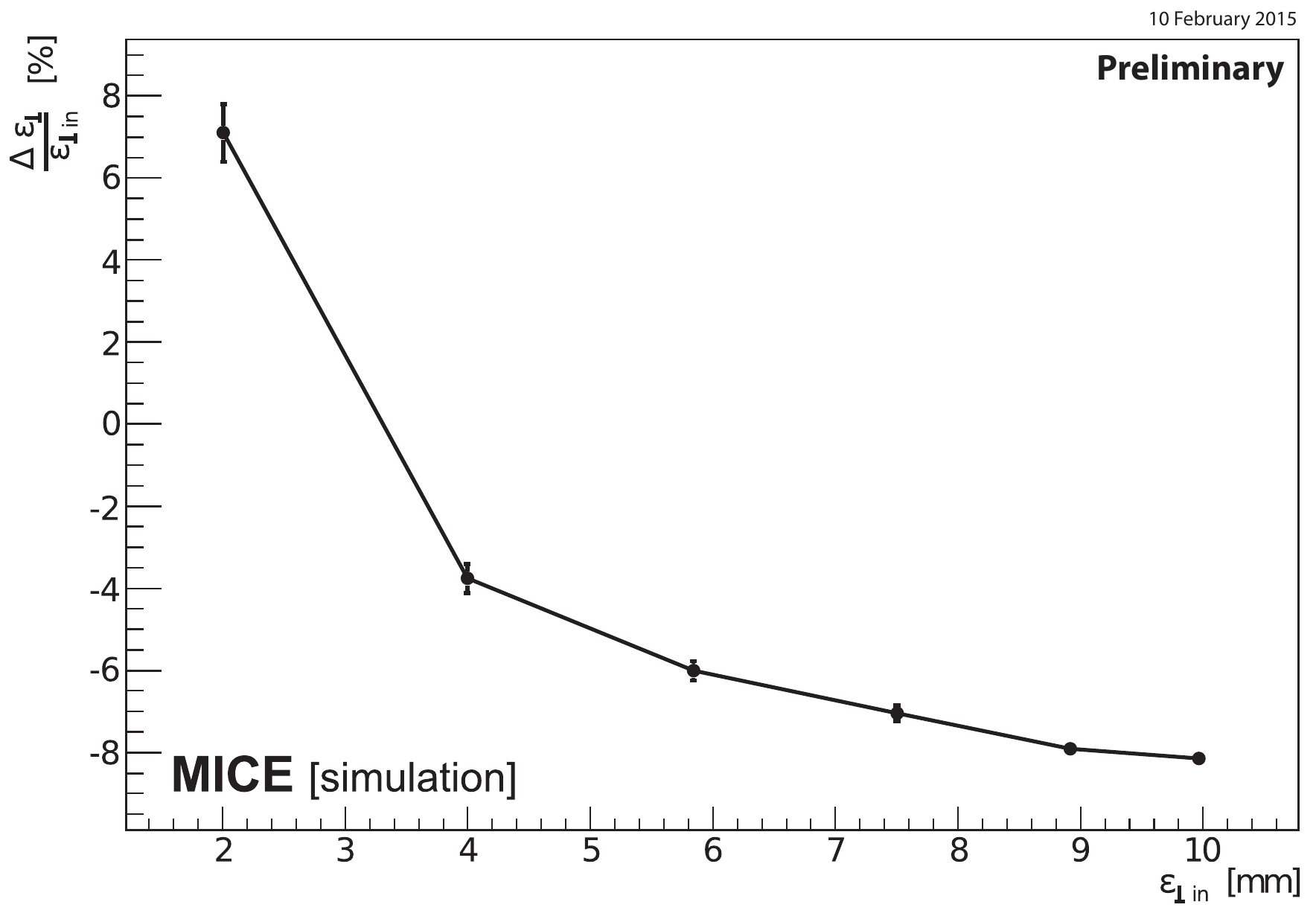}
\includegraphics[width=0.35\linewidth]{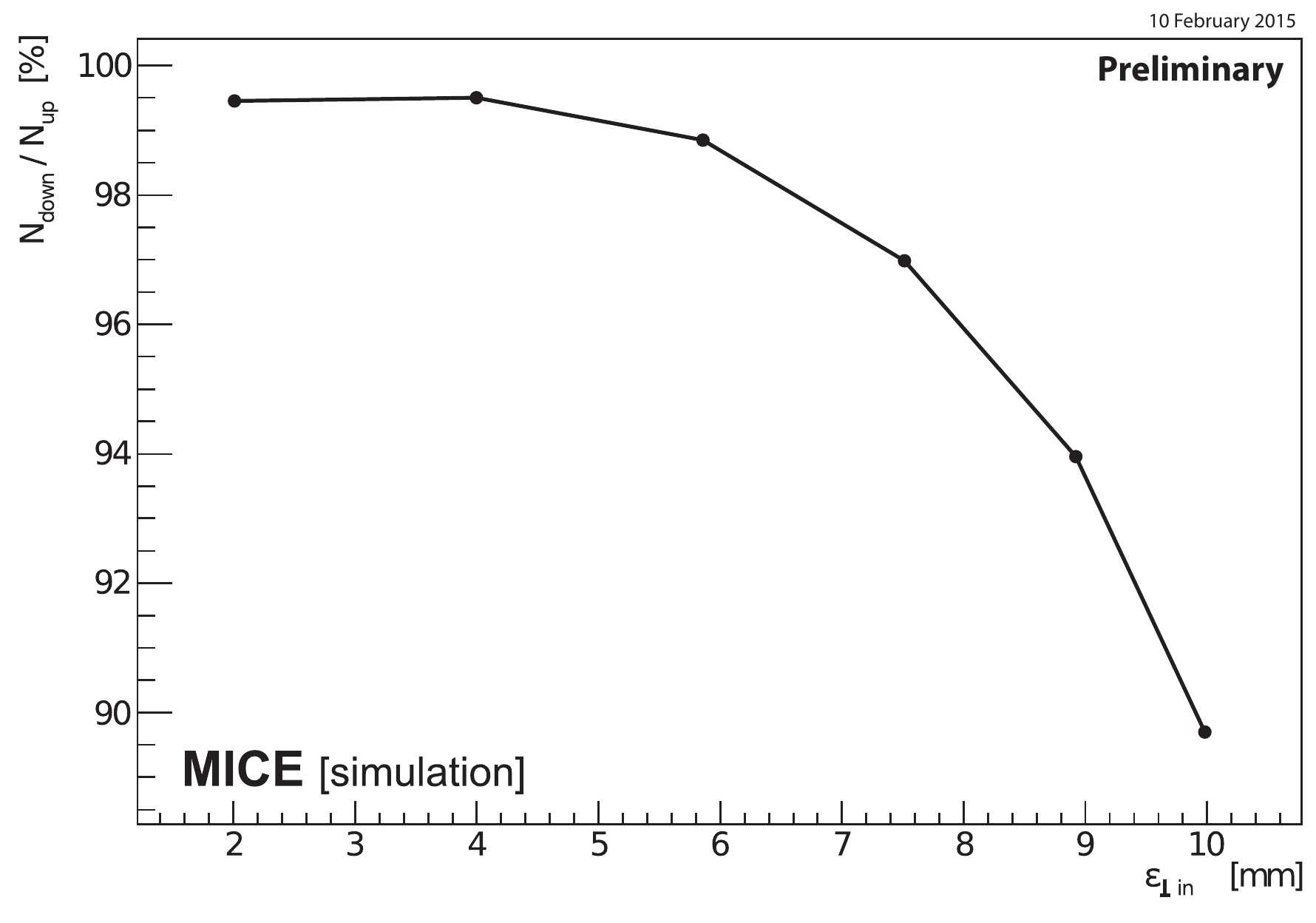}
\includegraphics[width=0.48\linewidth]{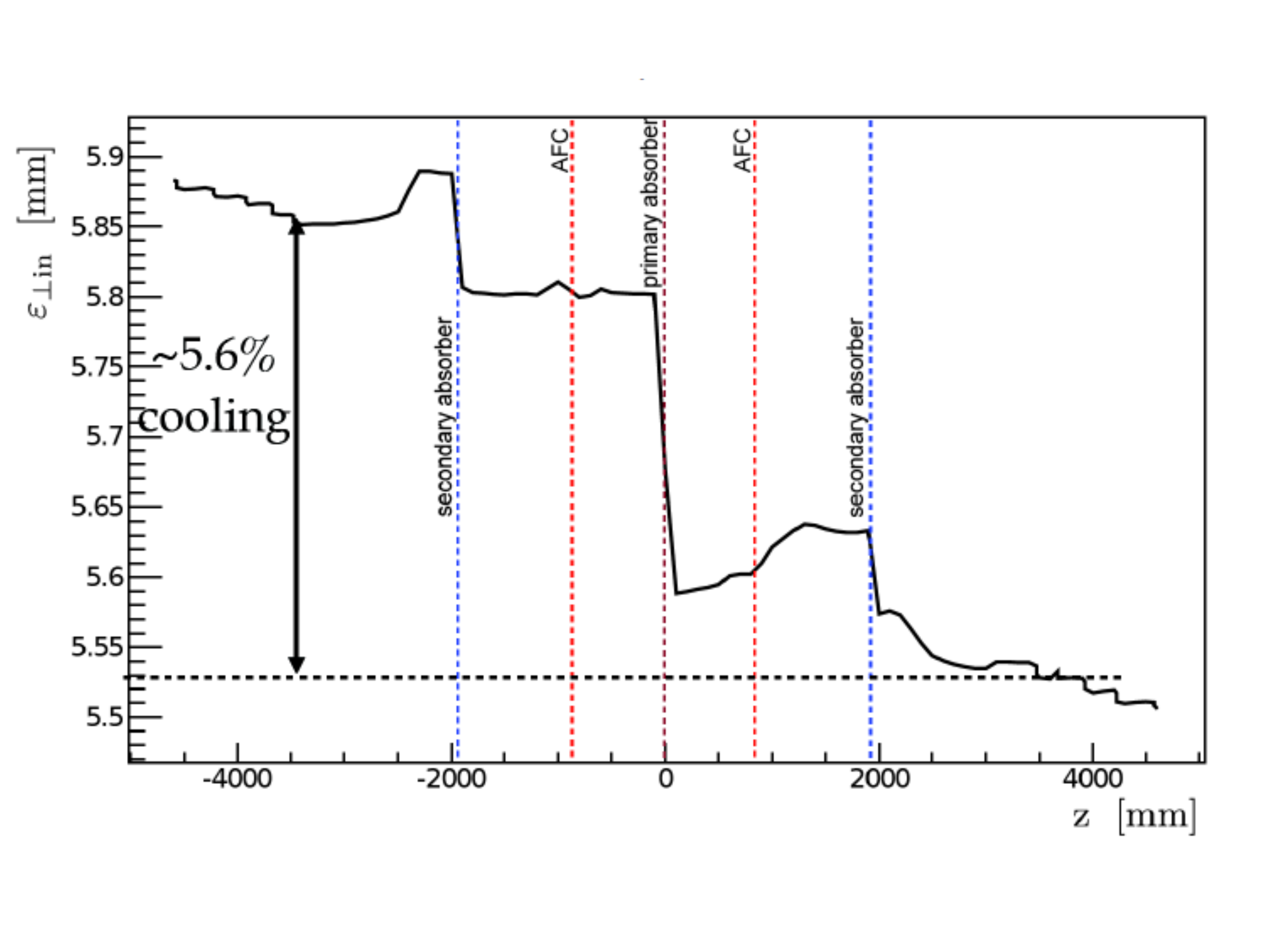}
\end{center}
\vskip -1.1cm
\caption{ Top-left panel: change in normalised 4D emittance as a function of
input emittance. Top-right: beam transmission 
as function of the input emittance. Bottom panel:
evolution of the 4D emittance in the ionization-colling demo lattice.
All plots are  for a $6 \pi \cdot$ mm, 200 MeV/c muon beam.}
\label{fig:dic}
\end{figure}
The initial baseline design with two RF-cavity-coupling-coil (RFCC) 
modules \footnote{each contained four 201 MHz RF cavities and one coupling-coil
(CC) solenoid}\cite{US2a}, providing re-acceleration between AFC modules, has been 
replaced by a simpler design with two single RF cavities, one central (65 mm) 
LiH absorber for cooling and two secondary focus-coil modules for 
beam focussing, as shown in
figure \ref{fig-mice}. 
In this way, the difficult task of integrating the CC with the RF cavities 
into a cryostat has been avoided, eliminating a high-risk schedule issue.
With this modified setup, a reduction of transverse emittance $\sim 5.6 \%$
with a transmission of $99 \%$ for a $6 \pi \cdot mm$ input emittance beam
(as respect to the baseline $10 \%$ reduction) is expected, as seen in  
figure \ref{fig:dic}. 
Data taking is foreseen to start in 2017. 

\section{Conclusions}
The fist step of MICE, devoted  to the characterization of 
 the incoming muon beam,
has been  accomplished, by the construction of the muon beamline 
and the PID detectors. Obtained detector performances meet the
requirements and a preliminary measure of emittance, with TOF only, has been 
realized. STEP IV for the study of the properties that
determine the cooling performance of a low energy muon beam, has just
started data taking in mid 2015. This step will involve the complex operation
of superconducting magnets of novel design. The next step, the demonstration 
of ionization cooling with re-acceleration, is scheduled to start in 2017.    


\begin{thebibliography}{99}
\bibitem{Koshkarev} Koshkarev, D. G., CERN/ISR-DI/74-62,1974; 
M. Bogomilov {\it et al.}, Phys. ReV. ST. Accel. Beams 17 (2014) 12,121002.
\bibitem{Bonesini} Bonesini, M. and Guglielmi, A., Phys.Rept. 433 (2006),65.
\bibitem{Ankenbrandt} C.M. Ankenbrandt {\it et al.}, Phys. Rev. ST Accel. 
Beams 2 (1999) 081001; 
R.Palmer {\it et al.}, Nucl. Phys. Proc. Suppl. 51A (1996), 61;
R. Palmer {\it et al.}, Phys. Rev. ST Accel. Beams 8 (2005) 061003.  
\bibitem{Choubey} Choubey S. {\it et al}, Int. Design Study for 
         the Neutrino Factory, IDS-NF-20,2011, arXiv:1112.2853.
\bibitem{Rees} Rees, G.H. and Kelliher, D.J.,
      Proceedings 46th ICFA Advanced Beam Dynamics Workshop HB2010,2010, 
      p. 54-56; Edgecock R., Int.J.Mod.Phys.A26 (2011) 1736-1743
\bibitem{Geer1} Geer S., LINAC10 Conference , Tsukuba, Japan, 2010, 
xarXiv:1202.2140.
\bibitem{srinsky} A.N. Skrinsky and V.V. Parkhomchuk, Sov. J. Part. Nucl. 12 (1981)  223
\bibitem{mice} G. Gregoire {\it et al.} {\bf [MICE coll]} , MICE Proposal to RAL, 2003. 
\bibitem{US2a}  S.Ozaki {\it et al.}, BNL-52623, June 2001 ; 
M.M. Alsharo'a {\it et al.}, Phys. Rev. ST. Accel. Beams 6,081001 (2003),  
arXiv:hep-ex/0207031;
R. Palmer et al., arXiv:0711.4275.
\bibitem{yordan} R. Bertoni {\it et al.}, Nucl. Instr. and Meth. A615 (2010) 14,
arXiv:001.4426.
\bibitem{ckv} L. Cremaldi {\it et al.}, IEEE Trans.  Nucl. Sci. 56 
 (2009) 1475.
\bibitem{tof2} R. Bertoni {\it et al.}, ``The construction of the MICE TOF2 
         detector'', MICE-NOTE-DET-286 (2010).  
\bibitem{bonesini2} M. Bonesini {\it et al.}, Nucl. Instr. and Meth. A693 (2012) 130.
\bibitem{kloe} A. Aloisio {\it et al.} {\bf [KLOE coll.]}, Nucl. Inst. and Meth. A 494 (2002) 326.
\bibitem{lietti} D.Lietti {\it et al.}, Nucl.Instr. and Meth. A604 (2009) 314;
R. Asfandiyarov {\bf [MICE coll]} , Nucl. Instr. and Meth. A732 (2013) 451.
\bibitem{Ellis} M. Ellis {\it et al.}, Nucl. Instr. and Meth A659 (2011) 136, 
arXiv:1005.3491.
\bibitem{bogo1} M. Bogomilov {\it et al.} {\bf [MICE coll.]} , JINST 7 (2012) P05009, arXiv:1203.4089
\bibitem{mark} D. Adams {\it et al.} {\bf [MICE coll.]}, Eur. Phys. J. C73 (2013) 2582, 
arXiv:1306.1509.
\bibitem{bonesini1} M. Bonesini JINST 9 (2014) 01, C01064.
\bibitem{pi-cont} M. Bogomilov {\it et al.} {\bf [MICE coll.]}, ``Pion  
contamination in the MICE
$\mu$  beam'', submitted to JINST.
\end{thebibliography}
\end{document}